\documentclass[doublecol]{epl2}
\usepackage{amssymb}
\usepackage{amsmath}
\usepackage{graphicx}
\usepackage{dcolumn}
\usepackage{bm}

\institute{School of Physics and Optoelectronic Technology, Dalian
University of Technology, Dalian 116024, P. R. China} \abstract{ A
general law is presented for (composite) quantum systems which
directly describes the time evolution of quantum states (with one or
both components) through an arbitrary noisy quantum channel. It is
shown that the time evolution of all quantum states through a
quantum channel can be completely captured by the evolution of a
single 'probe state'. Thus in order to grasp the information of the
final output states subject to a quantum channel, especially an
unknown one, it only requires quantum state tomography of a single
probe state, which dramatically simplifies the practical operations
in experiment. } 

\begin{document}

\title{Describing a Quantum Channel by State Tomography of a Single Probe State}
\author{Chang-shui Yu \thanks{%
quaninformation@sina.com; ycs@dlut.edu.cn}, He-shan Song}
\date{\today }
\maketitle

Quantum states are the basic carrier of quantum information [1]. The
core of all the quantum information processing (QIP) including
quantum communication [2] and quantum computation [3] is the
controlled time evolution of quantum state in essence [4]. However,
in realistic scenario, quantum states will be unavoidably and
greatly disturbed by the undesired coupling to the uncontrolled
degree of freedom usually termed as 'environment' and described as a
'quantum channel'. As a consequence, besides the state itself the
valuable properties of quantum states such as coherence [5,6],
entanglement [7,8] of composite systems and so on will be greatly
corrupted. The precise characterization of some properties of
quantum states usually largely relies on the evaluation of quantum
states, if these properties such as entanglement does not correspond
to a direct observable for a general unknown quantum state [9-11].
Furthermore, quantum channel is not restricted to the previous
interaction between the system and environment. It is a general
notion of any a input/output device governed by quantum mechanics
including the controlled interactions, for example, the dynamical
action of a quantum gate in a quantum computer etc [12]. Therefore,
it is of practical importance to precisely explore the time
evolution of quantum states on which a reliable QIP task depends.

In general cases, there is no direct way to evaluate the time
evolution of quantum states.  One has to begin with considering the
dynamics of 'system of interests + environment' governed by quantum
principle [13-16]. It is implied that the concrete description of
the quantum channel has been known by the quantum process tomography
that includes a series of quantum state tomography [17] and is
usually quite complex [3]. In experiment, the time evolution of
quantum states could be described by determining the initial and
final states in terms of quantum state tomography no matter whether
the quantum channel is known. However, it is a drawback that the
procedure needs to be repeated every time with different input
states chosen. In the present Letter, we provide a direct and
general scheme in terms of the evolution of a given probe state to
describe the evolution of quantum states of an arbitrary (composite)
quantum system which (the components of which) undergoes an
arbitrary (especially unknown) quantum channel.  The distinguished
advantages of our scheme are as follows: 1) Especially for unknown
quantum channels, it is only necessary to do state tomography of the
single probe state, stead of repeating the same procedure for
different input states or doing quantum process tomography (In other
words, it is not necessary to know the concrete description of
quantum channel). 2) The scheme can be directly applied to any
quantum mechanical input/output process. Thus all information of the
final states can be learned and the properties of interests such as
coherence or entanglement etc. can be obtained by a sequent simple
calculation [18,19].

Let us first consider an $(N\otimes N)$ -dimensional bipartite quantum
states $\rho _{0}$ which can be expanded in a representation spanned by
maximally entangled states given by
\begin{equation}
\left\vert \Phi _{j}\right\rangle =\frac{1}{\sqrt{N}}\sum%
\limits_{k=0}^{N-1}e^{i\frac{2j_{0}k\pi }{n}}\left\vert k\right\rangle
\left\vert k\oplus j_{1}\right\rangle ,j=Nj_{0}+j_{1},
\end{equation}%
where $j_{0},j_{1}=0,1,\cdots ,N-1,$ $\left\vert k\right\rangle $ is the
computational basis and '$\oplus $' denotes the addition modulo $N.$ Suppose
each subsystem of $\rho _{0}$ undergoes a quantum channel represented by \$$%
_{1}$ and \$$_{2}$, respectively, then the final state can be given by $\rho
_{f}=\left( \$_{1}\otimes \$_{2}\right) \rho _{0}/p$, where $p=$Tr$\left[
\left( \$_{1}\otimes \$_{2}\right) \rho _{0}\right] $ is the joint
probability for channels $\$_{1}\ $and $\$_{2}$ which corresponds to
non-trace-preserving channels [20]. In the representation of maximally
entangled states, $\rho _{f}$ can be expanded as%
\begin{equation}
\rho _{f}=\frac{1}{p}\sum\limits_{mn}\left( \$_{1}\otimes \mathbf{1}\right) %
\left[ \left\vert \Phi _{m}\right\rangle \left\langle \Phi _{m}\right\vert
\left( \mathbf{1}\otimes \$_{2}\right) \rho _{0}\left\vert \Phi
_{n}\right\rangle \left\langle \Phi _{n}\right\vert \right] ,
\end{equation}%
with $\mathbf{1}$ being the identity. Throughout the letter, we refer to $%
(N\times N)$ matrix $\psi $ (i.e. without $\left\vert {}\right\rangle $) as
the matrix notation of any pure state $\left\vert \psi \right\rangle
=\sum\limits_{i,j=0}^{N-1}a_{ij}\left\vert ij\right\rangle $ with matrix
elements $\left\langle i\right\vert \psi \left\vert j\right\rangle =a_{ij}$.
For a maximally entangled state $\left\vert \Phi _{m}\right\rangle $, it
hence follows that [21]
\begin{equation}
\left\vert \Phi _{m}\right\rangle =\left( \Phi _{m}P^{-1}\otimes \mathbf{1}%
\right) \left\vert P\right\rangle ,
\end{equation}%
where $\left\vert P\right\rangle =\sum\limits_{i,j=0}^{N-1}\tilde{a}%
_{ij}\left\vert ij\right\rangle $, called 'probe quantum state', is a
generic entangled pure state with full-rank $P$(which can be explicitly
written as
\begin{equation*}
P=\left(
\begin{array}{cccc}
\tilde{a}_{00} & \tilde{a}_{01} & \cdots & \tilde{a}_{0\left( N-1\right) }
\\
\tilde{a}_{10} & \tilde{a}_{11} & \cdots & \tilde{a}_{1\left( N-1\right) }
\\
\vdots & \vdots & \ddots & \vdots \\
\tilde{a}_{\left( N-1\right) 0} & \tilde{a}_{\left( N-1\right) 1} & \cdots &
\tilde{a}_{\left( N-1\right) \left( N-1\right) }%
\end{array}%
\right),
\end{equation*}
or can be directly obtained by the method provided above (3).) and $P^{-1}$
denotes the inverse matrix of $P$. $\Phi _{m}$ in (3) are simple unitary
transformations determined by (1). For example, for the state of a pair of
qubits, $\Phi _{m}$ corresponds to the three Pauli matrices and the
identity, respectively. Thus $\rho _{0}=\sum\limits_{k}p_{k}\left( \phi
_{k}P^{-1}\otimes \mathbf{1}\right) \left\vert P\right\rangle \left\langle
P\right\vert $ $\times \left( \left[ P^{\dag }\right] ^{-1}\phi _{k}^{\dag
}\otimes \mathbf{1}\right) $ for a potential decomposition $\{p_{k},\phi
_{k}\}$. Based on Jamio{\l }kowski isomorphism [22], (2) can be rewritten as

\begin{eqnarray}
\rho _{f} &=&\frac{1}{p}\sum\limits_{mnk}p_{k}\left\langle \Phi
_{m}\right\vert \left( \phi _{k}P^{-1}\otimes \mathbf{1}\right) \left[
\left( \mathbf{1}\otimes \$_{2}\right) \left\vert P\right\rangle
\left\langle P\right\vert \right]  \notag \\
&&\times \left( \left[ P^{\dag }\right] ^{-1}\phi _{k}^{\dag }\otimes
\mathbf{1}\right) \left\vert \Phi _{n}\right\rangle \left( \$_{1}\otimes
\mathbf{1}\right) \left[ \left\vert \Phi _{m}\right\rangle \left\langle \Phi
_{n}\right\vert \right]  \notag \\
&=&\frac{p_{I}p_{II}}{p}\sum\limits_{mn}\text{Tr}\left\{ S\left(
P^{-1}\otimes \Phi _{m}^{\ast }\right) \left[ \frac{\left( \mathbf{1}\otimes
\$_{2}\right) \left\vert P\right\rangle \left\langle P\right\vert }{p_{II}}%
\right] \right.  \notag \\
&&\times \left. \left( \left[ P^{\dag }\right] ^{-1}\otimes \Phi
_{n}^{T}\right) S\rho _{0}^{\ast }\right\} \left( \mathbf{1}\otimes \Phi
_{m}^{T}\left[ P^{T}\right] ^{-1}\right)  \notag \\
&&\times \left[ \frac{\left( \$_{1}\otimes \mathbf{1}\right) \left\vert
P\right\rangle \left\langle P\right\vert }{p_{I}}\right] \left( \mathbf{1}%
\otimes \left[ P^{\ast }\right] ^{-1}\Phi _{n}^{\ast }\right) ,
\end{eqnarray}%
where $S$ is the swapping operator defined as $S\left\vert j\right\rangle
\left\vert k\right\rangle =\left\vert k\right\rangle \left\vert
j\right\rangle $, '$T$' denotes transpose operation, $p_{I}=$Tr$\left[
\left( \$_{1}\otimes \mathbf{1}\right) \left\vert P\right\rangle
\left\langle P\right\vert \right] $, $p_{II}=$Tr$\left[ \left( \mathbf{1}%
\otimes \$_{2}\right) \left\vert P\right\rangle \left\langle P\right\vert %
\right] $ and $p$ is the normalization factor of (4). Once we select
one probe state $\left\vert P\right\rangle $, the evolution of $\rho
_{0}$ will completely be determined by the two output probe states
$\frac{\left( \$_{1}\otimes \mathbf{1}\right) \left\vert
P\right\rangle \left\langle P\right\vert }{p_{I}}$ and $\frac{\left(
\mathbf{1}\otimes \$_{2}\right) \left\vert P\right\rangle
\left\langle P\right\vert }{p_{II}}$ up to several fixed operations
which are independent of the initial input states and quantum
channels. In experiment, so long as one does state tomography of the
final states of the probe state corresponding to the two quantum
channels, one can directly obtain the final output state of
$\rho_0$. Thus (4) provides a general, direct and simple
relationship between the input state $\rho _{0}$ and the output
state $\rho _{f}$. Quantum channel's action on the initial quantum
state has been completely captured by a probe quantum state
$\left\vert P\right\rangle $ which is alternative depending on the
experimental convenience. The
maximally entangled state $\left\vert \Phi \right\rangle =\frac{1}{\sqrt{N}}%
\sum\limits_{k=0}^{N-1}\left\vert k\right\rangle \left\vert k\right\rangle $
is a special probe state with which (4) has a simple form by replacing $P$
by $\mathbf{1}$. Note that it is not necessary to employ the same probe
state $\left\vert P\right\rangle $ for both channels $\$_{1}$ and $\$_{2}$.

The direct relationship (4) can be reduced to describe the time evolution of
an $N$-dimensional quantum state $\varrho _{0}$ through a quantum channel $\$
$. To derive the explicit expression of the time evolution, we first
introduce an auxilliary $N$-dimensional quantum state $\sigma $ which is
completely a mathematical skill to double the orignal quantum state space.
Thus the bipartite joint quantum state can be written as $\rho _{0}^{\prime
}=\varrho _{0}\otimes \sigma $ and the corresponding final state $\rho
_{f}^{\prime }=\frac{1}{p_{s}}\$\varrho _{0}\otimes \sigma $ with $p_{s}=$Tr$%
\left( \$\varrho _{0}\right) $, which is equivalent to the case of one-sided
quantum channel in (4) and can be directly obtained by setting $\rho
_{0}=\rho _{0}^{\prime }$, $\$_{1}=\$$ and $\$_{2}=\mathbf{1}$. Hence the
output final state corresponding to $\varrho _{0}$ is given by $\varrho _{f}=
$ Tr$_{a}\rho _{f}^{\prime }$ \ with Tr$_{a}$ denoting trace over the
auxiliary system. It is obvious that the final state $\varrho _{f}$ is
independent of the auxilliary state $\sigma $. Repeating the same procedure
to (4), one can obtain a compact form of $\varrho _{f}$:%
\begin{eqnarray}
\varrho _{f} &=&\frac{p_{III}}{p_{s}}\sum\limits_{mn}\left\langle \Phi
_{m}\right\vert \frac{\left( \$\otimes \mathbf{1}\right) \left\vert
P\right\rangle \left\langle P\right\vert }{p_{III}}\left\vert \Phi
_{n}\right\rangle   \notag \\
&&\times \Phi _{m}P^{-1}\varrho _{0}\left( P^{-1}\right) ^{\dag }\Phi
_{n}^{\dag },
\end{eqnarray}%
where $p_{III}=$Tr$\left( \left( \$\otimes \mathbf{1}\right)
\left\vert P\right\rangle \left\langle P\right\vert \right) $. (5)
does not include the auxiliary state itself as expected, hence
$\sigma $ can be arbitrary quantum states and does not work in
practical experiment. The output state given in (5) is completely
determined by the one-sided quantum channel's actions on the
$(N\otimes N)$ -dimensional probe quantum states. (5) provides a
general input/output relationship for a given quantum state through
an arbitrary quantum channel.

In fact, (5) is a general law suitable for all quantum states no matter
whether the quantum systems under consideration are composite (including
multipartite quantum states) or not. However it is of the most importance
that the auxiliary Hilbert space introduced should be completely consistent
with that under consideration, unless the quantum channel is a nonlocal one
acting on all components of the composite system which can equivalently be
mapped to a single quantum system. For example, if we consider an $%
(N_{1}\otimes N_{2})$ -dimensional bipartite quantum state through a
one-sided or two-sided quantum channel, the auxiliary should also be an $%
(N_{1}\otimes N_{2})$ -dimensional bipartite quantum state. Thus, the probe
quantum state should be an $\left[ (N_{1}N_{2})\otimes (N_{1}N_{2})\right] $
-dimensional generic bipartite entangled state which is a quadripartite
quantum state in essence. However, if both components of the composite
system undergo a common nonlocal quantum channel, only a genuine bipartite
probe state will be enough.

Let us finally compare (4) and (5) briefly when we consider an $(N\otimes N)$
-dimensional bipartite quantum state. In (4), the probe states are two
separate generic $(N\otimes N)$ -dimensional entangled states. However, in
(5) one has to prepare a generic $(N^{2}\otimes N^{2})$ -dimensional
entangled state (a quadripartite entangled state) as the probe state which
is obviously not as easy to implement in experiment [23] as that in (4).
However, we again would like to emphasize that (5) is a universal law for
all quantum states.

In conclusion, we have presented a direct and general input/output law for
the time evolution of quantum states with the system or its components
through an arbitrary quantum channel. It has shown that quantum channel's
action on any quantum states can be completely captured by a preconditioned
probe quantum state. In other words, for any quantum channel, especially for
an unknown one, it is enough to only explore the time evolution of one of
the probe states instead of repeating the same procedure for every potential
initial input state. Thus the output probe state, just like a quantum gate,
directly relates the input and output states by which all information on the
final output states can be learned.

This work was supported by the National Natural Science Foundation
of China, under Grant No. 10805007 and No. 10875020, and the
Doctoral Startup Foundation of Liaoning Province.

\end{document}